\documentclass[prd, amsfonts, twocolumn, nofootinbib, showpacs]{revtex4}

\usepackage{graphicx}

\newcommand{\be}{\begin{equation}}
\newcommand{\ee}{\end{equation}}
\newcommand{\bea}{\begin{eqnarray}}
\newcommand{\eea}{\end{eqnarray}}

\newcommand{\hethree}{{\rm ^3He}}
\newcommand{\hefour}{{\rm ^4He}}

\newcommand{\lisix}{{\rm ^6Li}}
\newcommand{\liseven}{{\rm ^7Li}}

\newcommand{\gapp}{\mathrel{\raise.3ex\hbox{$>$}\mkern-14mu
               \lower0.6ex\hbox{$\sim$}}}
\newcommand{\lapp}{\mathrel{\raise.3ex\hbox{$<$}\mkern-14mu
               \lower0.6ex\hbox{$\sim$}}}

\pacs{26.35.+c, 98.80.Cq, 98.80.Ft}

\begin{document}

\title{Solving the cosmic lithium problems with primordial
 late-decaying particles}

\author{Daniel Cumberbatch$^1$, Kazuhide Ichikawa$^2$,  Masahiro
Kawasaki$^2$, Kazunori Kohri$^{3,4}$, Joseph Silk$^{1}$  and  Glenn
D. Starkman$^{1,5}$}

%\email{dtc@astro.ox.ac.uk}

\affiliation{
$^1$ Astrophysics Department,~University~of~Oxford,~Oxford,~OX1~3RH,~UK.\\
$^2$ Institute for Cosmic Ray Research, University of Tokyo, Kashiwa,
Chiba 277-8582, Japan,\\
$^3$ Physics~Department,~Lancaster~University~LA1~4YB,~UK,\\
$^4$ Harvard-Smithsonian Center for Astrophysics, 60 Garden Street,
Cambridge, MA 02138,USA,\\
$^5$ 
Department of Physics, Case Western Reserve University, Cleveland,
OH~~44106-7079. \\
}

\begin{abstract}
We investigate the modifications to predictions for the abundances of
light elements from standard Big-Bang nucleosynthesis when exotic
late-decaying particles with lifetimes exceeding $\sim1\,$sec are
prominent in the early Universe.  Utilising a model-independent
analysis of the properties of these long-lived particles, we identify
the parameter space associated with models that are consistent with
all observational data and hence resolve the much discussed
discrepancies between observations and theoretical predictions for the
abundances of $\liseven$ and $\lisix$.
\end{abstract}

\maketitle

%%%%%%%%%%%%%%%%%%%%%%%%%%%%%%%%%%%%%%%%%%%%%%%%%%%%%%%%%%%%%%%%%%%%%%
\section{Introduction} \label{sec:introduction}
%%%%%%%%%%%%%%%%%%%%%%%%%%%%%%%%%%%%%%%%%%%%%%%%%%%%%%%%%%%%%%%%%%%%%%

Standard Big-Bang nucleosynthesis (SBBN) is one of the most reliable
and farthest reaching probes of early Universe cosmology, being based
on the rigorously tested Standard Model of particle physics as well as
basic principles of nuclear physics.  Augmenting these principles with
experimental data concerning nuclear reactions, we can precisely
estimate the relative abundances of light elements (ALEs),
particularly D, $^3$He, $\hefour$, $\lisix$, and $\liseven$ (relative
to H) at the end of the ``first three minutes'' after the Big-Bang, as
a function of the baryon-to-photon ratio $\eta$.  Consequently, SBBN
has been instrumental in the realisation that baryonic matter
constitutes only a small proportion of the total energy density of the
Universe, hence providing further supporting evidence for the
existence of dark matter.  Utilising the value of $\eta$ measured by
the Wilkinson Microwave Anisotropy Probe (WMAP), $\eta^{\rm
WMAP}=(6.10\pm 0.21) \times 10^{-10}$ at the 68\%
C.L. ~\cite{Spergel:2006hy}, the majority of theoretical predictions
from SBBN are in excellent agreement with observational data.  This is
truly remarkable considering that the ALEs span many orders of
magnitude from $\hefour$/H$\sim0.08$ down to
$\liseven$/H$\sim10^{-11}$.

Despite the great success of SBBN,
it has been noted that the prediction for the ratio of $^7$Li to H,
$-9.476<{\rm log}_{10}(^{7}{\rm Li/H})<-9.322$
(obtained using $\eta^{\rm WMAP}$),
does {\it not} agree with current observations.
Recently Bonifacio {\it et al.}~\cite{Bonifacio:2006au} reported that
log$_{10}(^{7}$Li/H)$_{\rm obs} = -9.90~\pm~0.09$,
confirming the results of Ryan {\it et al.}~\cite{RBOFN}
log$_{10}(^{7}$Li/H)$_{\rm obs} = -9.91~\pm~0.10$.
Even the less stringent observational result $-9.63 \pm 0.06$
of Melendez and Ramirez (MR) \cite{Melendez:2004ni},
exceeds the SBBN prediction at 2\,$\sigma$.
This is the ``$^7$Li problem''.
It has been often argued that the $^7$Li abundance
would be smaller than the SBBN value due to depletion in
stars~\cite{Pinsonneault:2001ub,Korn:2006tv}.
However, a quite uniform $^7$Li abundance (the Spite plateau) observed in metal-poor stars
is somewhat difficult to explain with such stellar depletion and, moreover,
the recent detections by \cite{Asplund:2005yt} of more fragile isotope $^6$Li in some of these stars
provide evidence against stellar depletion.

The observations of \cite{Asplund:2005yt} that some metal-poor stars have the isotopic ratio $^6$Li/$^7$Li
of a few percent not only sharpened the $^7$Li problem but also
elucidated the ``$^6$Li problem". Since the $^6$Li abundance predicted in SBBN is
very small ($^6$Li/$^7$Li$\sim 3.3 \times 10^{-5}$ for $\eta^{\rm WMAP}$),
it is usually considered to be produced later through cosmic ray nucleosynthesis,
but it is also known that the conventional processes are not sufficient for the observed abundance.
This $^6$Li problem was exacerbated by \cite{Asplund:2005yt} who found a relatively
metallicity-independent abundance of $^6$Li which is in contrast to the prediction of the cosmic
ray synthesis scenario. In particular, a high abundance $^6$Li/$^7$Li = $0.046 \pm 0.022$ observed in
the very metal-poor star LP 815--43 with $[{\rm Fe/H}]= -2.74$, is very hard to obtain.
Unconventional scenarios to enhance $^6$Li are investigated, for example, in \cite{astroLi6prod}
but they cannot solve the $^7$Li problem in any case.

As has been discussed, the ``Lithium problems", too much $^7$Li and too little $^6$Li produced in
the standard scenario, do not have an astrophysical solution in a complete manner at present.\footnote{
Solutions by invoking not well measured reaction rates were investigated in
\cite{Coc:2003ce,Angulo:2005mi}, but without success at this stage. }
In addition, the recent observations seem to imply that stellar depletion of $^7$Li  is limited and that
the metallicity dependence of $^6$Li is only modest.
These facts lead us to seek a solution to the  Lithium problems by
incorporating particle physics  beyond the standard model.
Specifically, we reinvestigate the effects on BBN predictions  of
late-decaying particles (LDPs)  possessing a finite hadronic branching
ratio.  In the mid-eighties, Dimopolous, Esmailzadeh, Hall
and Starkman (DEHS)  showed that the mixed hadronic and
electromagnetic decays of a  massive particle at $t\sim10^5$s could
reproduce the  ratios of light elements
\cite{Dimopoulos:1987fz,Dimopoulos:1988ue} as then measured.  In LDP
nucleosynthesis, the products of particle decays   scatter off the
SBBN-produced light elements modifying the ALEs.  Potentially, the
modifications in the  ALEs could eliminate the existing
inconsistencies with current observations, since they predicted that a
signature of such decays  would be an anomalously high $^6$Li to $^7$Li
ratio.

LDPs, which we call here $X$, appear in widely considered extensions
to the standard model.  For example, they are realised
~\cite{Jedamzik:2004er,Kawasaki:2004yh,Kawasaki:2004qu,%%
Kohri:2005wn,Jedamzik-Li,Jedamzik:2006xz} in supergravity models where
the next-to lightest sparticle (NLSP) decays into the lightest
sparticle (LSP) with an extremely long lifetime (typically exceeding
$\sim1$\,sec), owing to the Planck-mass suppression of its
interactions~\cite{weinberg:1982}.  In these theories the most
favoured candidate for $X$ would be a gravitino with a neutralino
LSP~\cite{Kohri:2005wn}, or a neutralino, a stau and a sneutrino as
NLSPs with a gravitino
LSP~\cite{others, Ellis:2003dn,Feng:2004zu,Roszkowski:2004jd,Cerdeno:2005eu,%%
Steffen:2006hw,Kawasaki:2007xb,Kanzaki:2006hm}.  These are but a few of the plethora of subtly
varying possibilities.  Many have been discussed and their effects
rigorously investigated in studies conducted in the
1980's~\cite{BBNwX_OLD_80s,Dimopoulos:1987fz,Dimopoulos:1988ue},
1990's~\cite{BBNwX_OLD_90s}, right up to recent
times~\cite{BBNwX_OLD_00s,Kawasaki:2004yh,Jedamzik:2006xz}, with
significant improvements being made at each stage (see also a text
book~\cite{Khlopov:1999rs} and references therein, and a recent
partial reconfirmation in Ref.~\cite{Kusakabe:2006hc}).  Given the
excellent agreement between SBBN and the measured (or inferred)
abundances of D, $\hethree$ and $\hefour$, the properties of LDPs are
heavily constrained.  Such constraints are valuable to theories beyond
the standard model involving LDPs that are massive and
weakly-interacting, which are difficult to study in collider
experiments.

The purpose of this paper is to utilise the state-of-the-art model-independent
analysis of LDP nucleosynthesis performed by Kawasaki, Kohri and Moroi
(KKM) \cite{Kawasaki:2004yh,Kawasaki:2004qu} (incorporating important
improvements of the original DEHS analysis),
to identify the parameter space that is consistent with current
observations.  That is, we determine the range of values such as
hadronic branching ratio $B_h$, the lifetime $\tau_X$, and the
primordial energy density of $X$ particles divided by primordial
entropy density $Y_X$ that solves the Lithium problems while leaving the
abundances of H, D, $\hethree$ and $\hefour$ in agreement with
observations.  In addition we will reconfirm the original prediction
of DEHS that a signature of the model is a high $\lisix/\liseven$
ratio, and note that this prediction is now in agreement with
observational data.  Fundamental particle physics models with LDPs
would then need to lie {\it within} this allowed region of parameter
space in order to solve the Lithium problems.  Alternately, they might
satisfy limits which have been presented elsewhere to ensure that they
did not significantly change the ALEs, and look elsewhere for a
solution of the Lithium problems.

In Sec.~\ref{sec:reaction}, we describe some key reactions in solving the Lithium problems.
We carefully discuss how we can
reduce uncertainties with regards the non-thermally produced abundance of $^{7}$Li
and $^{7}$Be resulting from the $\alpha$--$\alpha$ collisions and make conservative predictions for the
corresponding effects  on the $^{7}$Li abundance.
In Sec.~\ref{sec:observation}, we summarise recent observational data which we try to
explain by BBN with LDPs. We present our results in Sec.~\ref{sec:result} and comment
on the solution when some stellar depletion takes place in Sec.~\ref{sec:depletion}.
We discuss our results in Sec.~\ref{sec:discussion}.

%%%%%%%%%%%%%%%%%%%%%%%%%%%%%%%%%%%%%%%%%%%%%%%%%%%%%%%%%%%%%%%%%%%%%%
\section{Reactions} \label{sec:reaction}
%%%%%%%%%%%%%%%%%%%%%%%%%%%%%%%%%%%%%%%%%%%%%%%%%%%%%%%%%%%%%%%%%%%%%%

The reaction which is most significant in reducing the net $^{7}$Li
abundance is the neutron capture of $^{7}$Be, $^7\rm{Be} + n \to
{^7}\rm{Li}+p$, where the neutron is non-thermally produced in the
hadronic shower resulting from the hadronic decay of a LDP.
Subsequently, $^{7}$Li can be destroyed by thermal protons through the
process $^7\rm{Li} + p \to 2\,{^4}\rm{He}$.  For relatively high
baryon to photon ratios, $\eta \gtrsim 3\times 10^{-10}$, the cosmic
$^{7}$Li abundance at much later times is mainly generated through
electron capture by primordial $^{7}$Be.  This is because primordial
$^{7}$Li (but not $^{7}$Be) is destroyed by the above thermal process
involving $p$ capture.  However, if large numbers of the non-thermal
neutrons are emitted (e.g. by LDP decay) at the appropriate time, then
the primordial $^{7}$Be abundance can be converted to $^7$Li
sufficiently early to be destroyed through thermal proton capture.
This mechanism was originally identified by Jedamzik in his pioneering
work~\cite{Jedamzik:2004er} for the first time and studied further in
detail
by~\cite{Kohri:2005wn,Jedamzik-Li,Jedamzik:2006xz,Jedamzik:2007cp}
(For other mechanisms to reduce the $^{7}$Li abundance beyond the
standard model, see also
\cite{ichikawa-Li7,Kohri:2006cn,Cyburt:2006uv,Bird:2007ge,Jittoh:2007fr,
Jedamzik:2007cp}).

The competing constraints on the properties of LDPs ultimately come from
preventing the overproduction of D and $^{6}$Li.
These elements are also  non-thermally produced
during the evolution of the LDP hadronic shower.
D is directly produced through the spallation of $^{4}$He by energetic nucleons,
while $^{6}$Li is mainly produced through the scattering of energetic (shower)
T and $^{3}$He off the background $^{4}$He~\cite{Dimopoulos:1987fz,%
Kawasaki:2004yh,Jedamzik:2004ip}.  Such energetic T and $^{3}$He are
produced  through the destruction of the background $^{4}$He within
the LDP hadronic shower.

Both $^{7}$Li and $^{7}$Be can also be non-thermally
produced in a manner similar to $^{6}$Li~\cite{Dimopoulos:1987fz}.
If $^{4}$He are produced during the evolution of the LDP hadronic shower
through $p/n$ +  $^{4}$He $\to$ $p/n$ +  $^{4}$He +$\pi$\,s
(due to energetic shower nucleons),
then energetic $^{4}$He can collide with background $^{4}$He
and produce $^{7}$Li and $^{7}$Be (and also $^{6}$Li,
although this is subdominant compared to
the T\,-\,$^{4}$He, and $^{3}$He\,-\,$^{4}$He processes \cite{Dimopoulos:1987fz,Kawasaki:2004yh,Kawasaki:2004qu}).
These non-thermal production mechanisms
were studied in detail in~\cite{Kawasaki:2004yh,Kawasaki:2004qu},
in order to constrain the properties of the LDPs,
and also applied to solve the $^{7}$Li problem
with some audacious approximations in~\cite{Kohri:2005wn}.
However, only if we adopt milder observational constraints on $^{7}$Li
(as in~\cite{Kawasaki:2004yh,Kawasaki:2004qu}),
can we ignore the intricacies of these production processes.

Unfortunately, there is a severe lack of experimental data
on the  energy distribution  of the $^{4}$He
in the final state  of  $\alpha$--$\alpha$  inelastic  scattering
in the relevant energy regime.
This  energy distribution  is essential
in order to accurately compute  the  abundance of
non-thermally produced $^{7}$Li  and $^{7}$Be.
By making reasonable approximations for the energy  of the  scattered $^{4}$He,
inferred from experimental  data on similar processes
involving collisions of energetic heavy ions,
and utilising the theoretical properties of quantum chromodynamics  (QCD),
some  of  the  current  authors  have  investigated  these  non-thermal
processes~\cite{Kawasaki:2004qu,Kohri:2005wn}.
The constraints  on  the abundance  of  hadronically-decaying LDPs
from predictions of the abundance of $^{7}$Li  and $^{7}$Be
were not significantly affected by these  approximations,
since the authors considered relatively generous observational constraints.

However, for our purposes it is essential
to more precisely calculate the abundances of these elements.
In particular,
a slight overestimation of the production of  $^{7}$Li and $^{7}$Be
may counterbalance their depletion by neutron capture,
resulting in a worse fit to the more stringent data now available.
Therefore we must reconsider the above ambiguities
in order to better estimate the non-thermally produced
$^{7}$Li and $^{7}$Be abundances.

In the previous treatment~\cite{Kawasaki:2004qu,Kohri:2005wn}, there
was a tendency for the estimated energy of the final state $^{4}$He
to be larger than the equipartition distribution  in the
center-of-mass (CM) system.   This was because the authors used
theoretical properties of quantum chromodynamics (QCD) to extrapolate
results from high-energy experiments to lower energies  until the
extrapolation became kinematically inconsistent.  However, whenever we
include non-standard processes,  such as the aforementioned
non-thermal $\alpha$--$\alpha$ collisions,  into the standard
calculations,  it would be better to adopt most conservative
approximations  that does not result in such an overestimate of the
kinetic energy  of the final state $^{4}$He.  Hence, we conservatively
chose a (smaller) value of the $^4$He kinetic energies  between the
experimentally-suggested QCD prediction~\cite{Moroi-suggestion} and
the equipartition value in the CM system.  For non-relativistic
$\alpha$--$\alpha$ collisions this tends to give smaller energies for the
scattered $^{4}$He, reducing the resulting abundance of Li and Be.
To understand the significance of these processes,  we also
investigated scenarios without the $\alpha$--$\alpha$ collisions.

%%%%%%%%%%%%%%%%%%%%%%%%%%%%%%%%%%%%%%%%%%%%%%%%%%%%%%%%%%%%%%%%%%%%%%
\section{Observational constraints on light element abundances} \label{sec:observation}
%%%%%%%%%%%%%%%%%%%%%%%%%%%%%%%%%%%%%%%%%%%%%%%%%%%%%%%%%%%%%%%%%%%%%%

Our theoretical ALEs must be compared against the
observational constraints on the abundances of
D, $^{4}$He, $^{6}$Li and $^{7}$Li.
The errors presented here are 1\,$\sigma$ errors unless otherwise stated.
The subscripts ``p'' and ``obs''
refer to the primordial and observational values, respectively.
The abundance of $^{3}$He does not play any significant role in our conclusions.

With regard to the mass fraction of $^{4}$He, recently two groups reported new
values of $Y_{\rm p}$~\cite{Peimbert:2007vm,Izotov:2007ed} by adopting
quite new $^{4}$He-emissivity data~\cite{Porter:2005vf}.
Izotov, Thuan and Stasinska reported two values,
$Y_{\rm p}$(IZS1) = $0.2472 \pm 0.0012$
and $Y_{\rm p}$(IZS2) = $0.2516 \pm 0.0011$~\cite{Izotov:2007ed}
by using old and new $^{4}$He-emissivity data, respectively.
Note that $Y_{\rm p}$(IZS2) at face value excludes the SBBN prediction ($\simeq 0.2484$)
even with various theoretical errors ($\simeq 0.0004$).
We artificially incorporate a larger error into the value of $Y_{\rm p}$(IZS2),
to investigate conservative bounds, and  call it ``IZS3'',
$Y_{\rm p}$(IZS3) = $0.2516 \pm 0.0040$,
where the larger error $0.0040$
was adopted according to a discussion in~\cite{Fukugita:2006xy}.
The possibility of these kinds of large
errors was also discussed in~\cite{Peimbert:2007vm}.

We adopt the following  two  deuterium abundances, Low (D/H)$_{\rm
obs} = (2.82 \pm 0.26) \times 10^{-5}$ as a most-recently reported
value of the weighted mean~\cite{O'Meara:2006mj},   and the more
conservative value,  High (D/H)$_{\rm obs} = (3.98^{+0.59}_{-0.67})
\times 10^{-5}$~\cite{Burles:1997fa}.

As was discussed in Sec.~\ref{sec:introduction}, we will compare both milder and more
stringent constraints on $^{7}$Li/H: log$_{10}(^{7}$Li/H)$_{\rm
obs}=-9.63~\pm~0.06$ (MR) ~\cite{Melendez:2004ni}, and
log$_{10}(^{7}$Li/H)$_{\rm obs}=-9.90~\pm~0.09$ (Bonifacio et al.)
\cite{Bonifacio:2006au,RBOFN}.

For the $^{6}$Li abundance, we adopt $^{6}$Li/$^{7}$Li = $0.046 \pm 0.022$,
which was recently observed in the very metal-poor star LP 815-43 with
$[\rm{Fe/H}] = -2.74$~\cite{Asplund:2005yt}.
Again note that the value of $^{6}$Li/$^{7}$Li calculated in SBBN
($\sim 3.3 \times 10^{-5}$) does not agree with this constraint.

%%%%%%%%%%%%%%%%%%%%%%%%%%%%%%%%%%%%%%%%%%%%%%%%%%%%%%%%%%%%%%%%%%%%%%
\begin{figure}
     \begin{center}
     \includegraphics[width=80mm,clip,keepaspectratio]{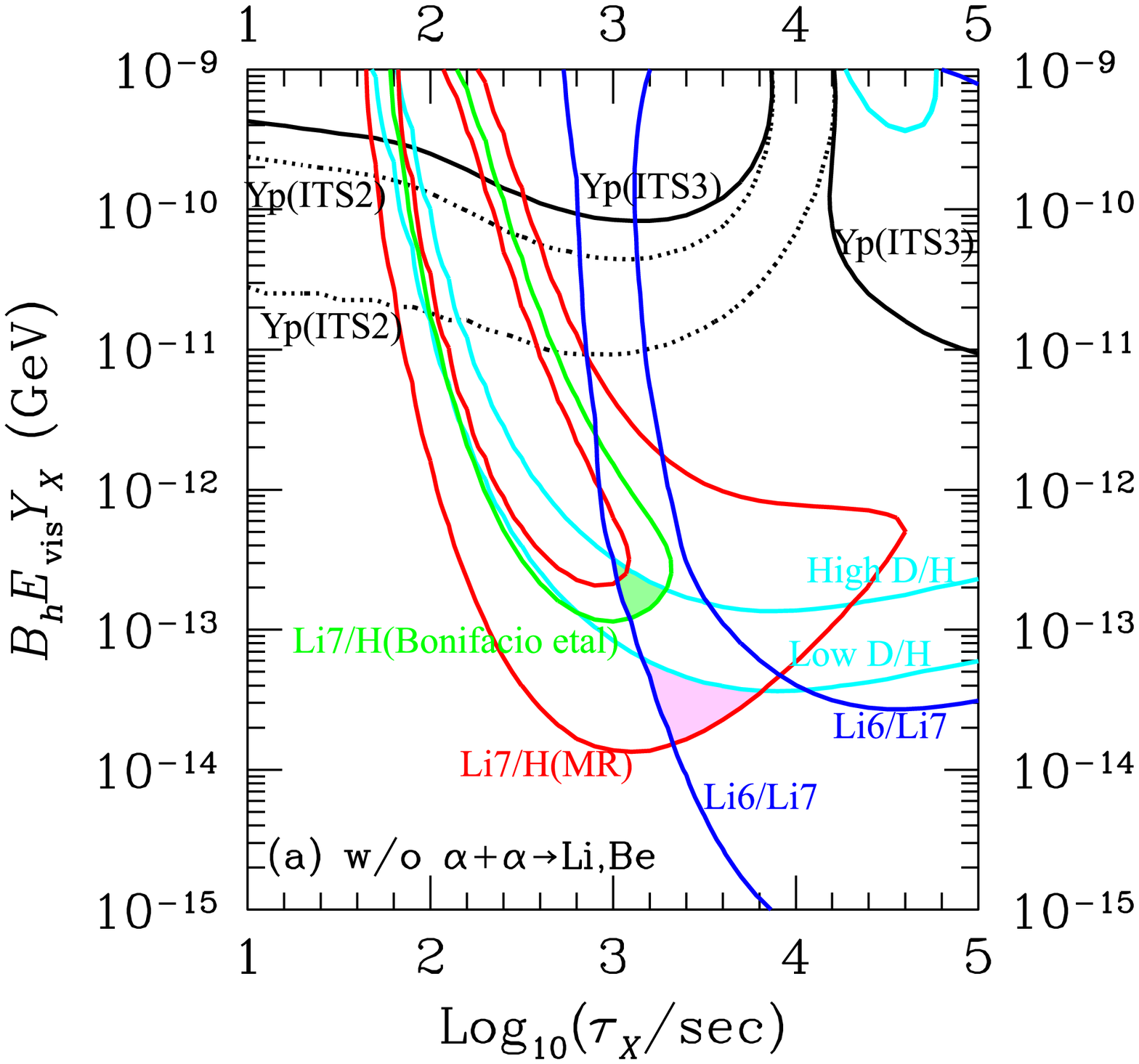}

%    \vspace{0.3cm}

     \includegraphics[width=80mm,clip,keepaspectratio]{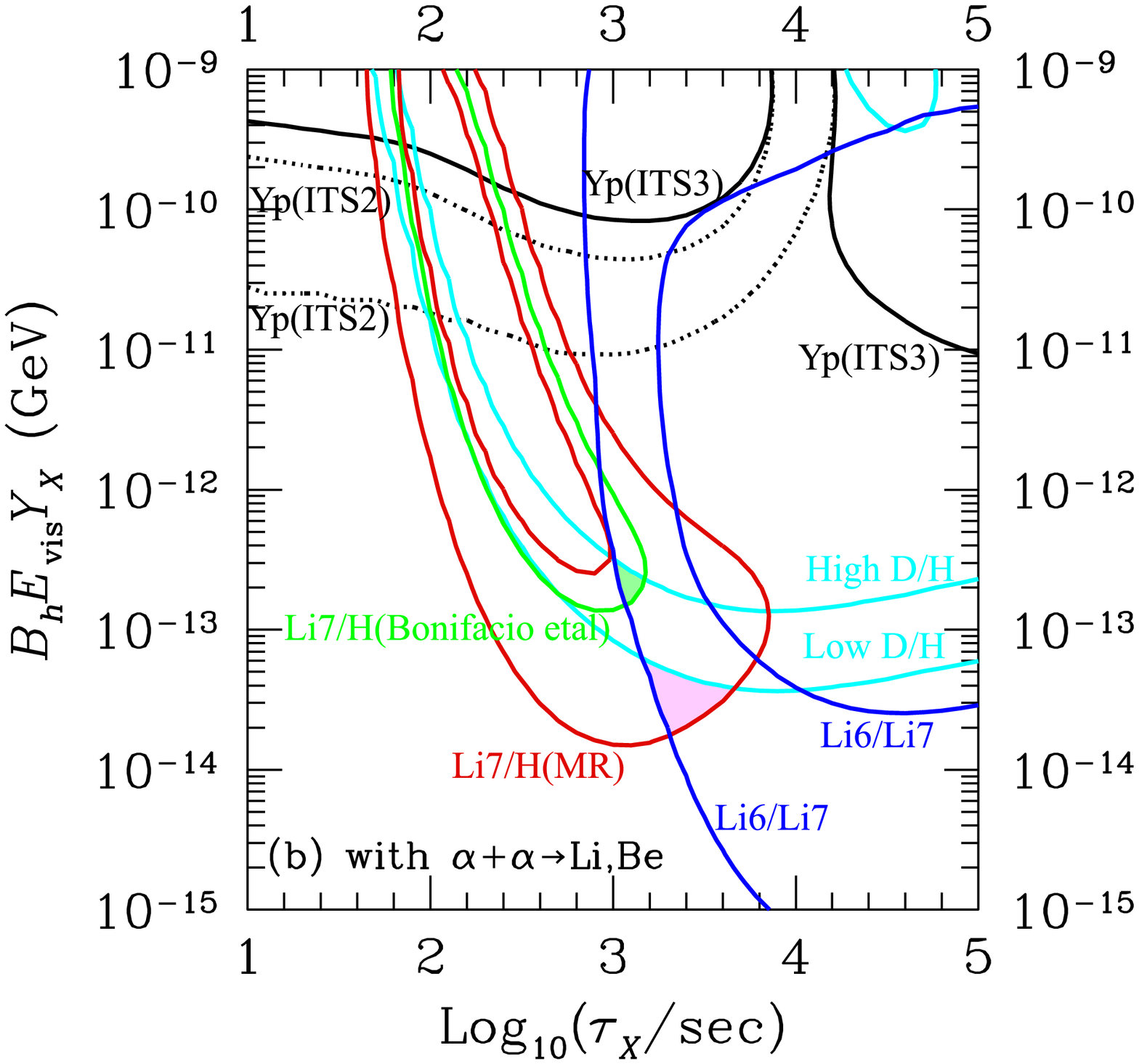}
     \end{center}

     \vspace{-0.5cm}

     \caption{Bounds on  $B_h E_{\rm vis}Y_X$ as a function of $\tau_X$.
     In the panel (a), $\alpha$--$\alpha$ collisions are excluded and in (b),
     $\alpha$--$\alpha$ collisions are included.
     In each panel,
     the lower (upper) shaded region  indicates the parameter space
     which is consistent with all observations when we adopt Low (D/H)
     and $^{7}$Li/H (MR) (High (D/H) and $^{7}$Li/H (Bonifacio et al.)
     ). The element's name is written beside
     each contour. }
     \label{fig:bounds}

     \vspace{-0.3cm}

\end{figure}
%%%%%%%%%%%%%%%%%%%%%%%%%%%%%%%%%%%%%%%%%%%%%%%%%%%%%%%%%%%%%%%%%%%%%%

%%%%%%%%%%%%%%%%%%%%%%%%%%%%%%%%%%%%%%%%%%%%%%%%%%%%%%%%%%%%%%%%%%%%%%
\section{Results} \label{sec:result}
%%%%%%%%%%%%%%%%%%%%%%%%%%%%%%%%%%%%%%%%%%%%%%%%%%%%%%%%%%%%%%%%%%%%%%

Let us first summarise the basic framework of our study.  We describe
the properties of the LDP, which we call $X$,  using only the following generalised
parameters:  $E_{\rm vis}$, the (averaged) energy emitted in the form of visible
particles (an invisible particle may also be emitted in some cases);
$\tau_X$, the lifetime of $X$;  $B_{\rm h}$, the branching ratio for
$X$ decay channels directly resulting in hadron production; the
primordial abundance of $X$,  which we parameterise using the ``yield
variable'' $Y_X \equiv n_X/s$,  which is defined at a cosmic time $t\ll \tau_X$.
Here, $n_X$ is the number density of $X$  while $s$ is the
total entropy density.  We know that the influence of unobservable
decay products (e.g. neutrinos)  has a negligible effect in our
calculations (see \cite{Kawasaki:1994bs,Kanzaki:2006hm,Kanzaki:2007pd}
and references therein).

%%%%%%%%%%%%%%%%%%%%%%%%%%%%%%%%%%%%%%%

In our analysis, we calculate the primordial ALEs  for a variety of
different combinations of the above LDP model parameters,  taking account of
dissociation processes  induced by the additional hadronic and
electromagnetic interactions  resulting from $X$ decays, as discussed
in \cite{Kawasaki:2004yh,Kawasaki:2004qu},  with some important
modifications discussed in Sec.~\ref{sec:reaction}.
We assume that two jets are produced in each hadronic decay of $X$,
each with an energy $E_{\rm jet}=m_X/2$,
with the hadronic branching ratio of $X$ equal to $B_{\rm h}$.
Here we set the mass of $X$ to $m_X$ = 1\,TeV
and the energy converted into visible particles to
$E_{\rm vis} = 2E_{\rm jet} = m_{X}$.
However, we conveniently find that
even if we change $m_X$, $E_{\rm vis}$ or $B_{\rm h}$,
the constraints on $B_{\rm h}E_{\rm vis}Y_X$ are not significantly altered for $\tau_X \lesssim 10^6$\, sec.
By comparing the results with the observations,
we derived stringent constraints on both $B_{\rm h}E_{\rm vis}Y_X$ and $\tau_X$.

The shaded region indicates the part of parameter space that is
consistent with observational 2$\sigma$ constraints  on the abundances of
$\hefour$(ITS3), D, $^{7}$Li and $^{6}$Li,  where in (a) we omit the
$\alpha$--$\alpha$ collisions as discussed in Sec.~\ref{sec:reaction},  whereas in (b) we
include the $\alpha$--$\alpha$ collisions.  Two combinations of D and
$^{7}$Li constraints are displayed: one using  Low (D/H) and
$^{7}$Li/H (MR), and a  second using High (D/H) and $^{7}$Li/H
(Bonifacio et al.).  Note that only {\it upper bounds} on $B_{\rm
h}E_{\rm vis}Y_{X}$  are provided by the $Y_{\rm p}$(ITS3) and D/H
contours.  We also note that, even if we adopted other observational
bounds on $^4$He,  such as ITS1 and the values published
in~\cite{Peimbert:2007vm,Fukugita:2006xy},  our results on the
consistent region of parameter space would not be affected.

In {\it both} Fig.\ref{fig:bounds} (a) and (b)
(i.e. with and without the incorporation of $\alpha$--$\alpha$ scattering),
we identify a region of parameter space that agrees with all observational constraints,
{\it including} those relating to $^{7}$Li/H and $^6$Li/H.
Even if we adopted the more stringent constraints claimed by Bonifacio {\it et al.},
there still remains such an allowed region of parameter space
{\it if} we adopt the High(D/H) result.
Any differences in these observationally-consistent regions of parameter space
in figures (a) and (b) should be interpreted as
due to theoretical uncertainties in the hadron shower physics.
Fortunately, such differences are minute.

The $^{6}$Li/$^{7}$Li contour located at smaller $\tau_X$
corresponds to the 2$\sigma$ lower bound  $^{6}$Li/$^{7}$Li\,$>$\,0.002.
It has the unusual feature of being nearly perpendicular to the $\tau_X$-axis.
This is because the  non-thermally produced $^{6}$Li  is uniformly destroyed
by  the standard thermal process $^{6}$Li($p$, $^{3}$He)$^{4}$He,
whose reaction rate is a steeply falling function of cosmic temperature.

If we  adopted a less-stringent  lower  bound,
say $^{6}$Li/$^{7}$Li $\gtrsim  6  \times 10^{-5}$
(note  that SBBN predicts $^{6}$Li/$^{7}$Li  $\simeq 3.3\times 10^{-5}$
for $\eta = 6.1\times 10^{-10}$),
it can explain $Y_{\rm p}$(ITS2).
In turn we may say that we require models
where $^{6}$Li/$^{7}$Li $\gtrsim  6  \times 10^{-5}$
in order to resolve the discrepancy
between the $Y_{\rm p}$(ITS2) measurement and the corresponding SBBN prediction.

%%%%%%%%%%%%%%%%%%%%%%%%%%%%%%%%%%%%%%%%%%%%%%%%%%%%%%%%%%%%%%%%%%%%%%
\section{Stellar Depletion} \label{sec:depletion}
%%%%%%%%%%%%%%%%%%%%%%%%%%%%%%%%%%%%%%%%%%%%%%%%%%%%%%%%%%%%%%%%%%%%%%

There is a possibility of late-time destruction of $^{6}$Li and $^{7}$Li in stars.
Although we have argued in Sec.~\ref{sec:introduction} that the detections of $^6$Li in
\cite{Asplund:2005yt} show there is no stellar depletion, it can change for BBN with LDPs
since $^6$Li can be produced so much relative to $^7$Li that the observed ratio $^6$Li/$^7$Li
is reproduced even though the depletion takes place.
In fact, we find an interesting solution for which the constraints on the abundance of $^7$Li
claimed by Bonifacio {\it et al.} and Ryan {\it et al.}
are compatible with the Deuterium abundance constraint,
Low (D/H), with some stellar depletion.

According to \cite{Pinsonneault:1998nf,Pinsonneault:2001ub},
rotational mixing within halo stars  results in the depletion of
$^{7}$Li, which can be parameterised by a depletion factor $D_{7}$.
The currently-observed $^{7}$Li abundance  is then related to its
primordial value by  $\log_{10}$($^{7}$Li/H)$_{\rm p}$ =
$\log_{10}(^{7}$Li/H)$_{\rm obs} + D_{7}$.  Because $^{6}$Li is more
easily dissociated than $^{7}$Li,  the depletion factor associated
with $^{6}$Li,  $D_{6} \simeq 2.5D_{7}$~\cite{Pinsonneault:2001ub}.
Thus ($^{6}$Li/$^{7}$Li)$_{\rm p}$ = ($^{6}$Li/$^{7}$Li)$_{\rm obs}
\times 10^{1.5 D_{7}}$.

\begin{figure}
     \begin{center}
     \includegraphics[width=80mm,clip,keepaspectratio]{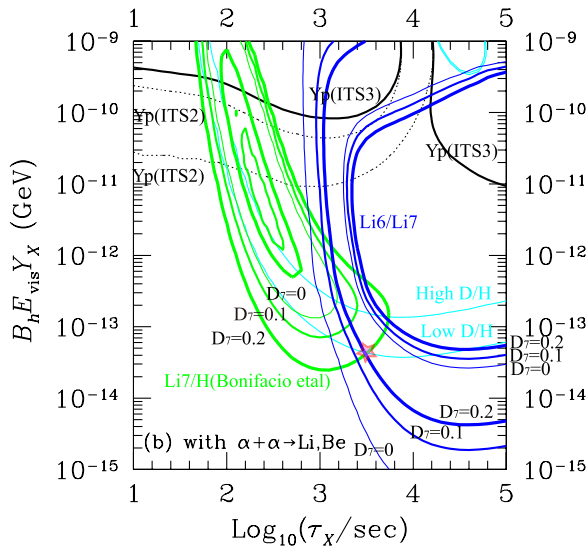}
     \end{center}

     \vspace{-1.3cm}

     \caption{
     Results assuming stellar depletion.  Here we have adopted the
     depletion factor $D_{7}$ = 0, 0.1, and 0.2 (thin line, moderate line, and thick line).
      We see that the most
     severe bounds on $^{7}$Li  agrees with  the theory with $D_{7}$ =
     0.2. The star symbol means the allowed region. Here we have
     included $\alpha+\alpha \to$ Li and Be processes.}

     \vspace{-0.5cm}

     \label{fig:D7}
\end{figure}

In Fig.~\ref{fig:D7}, we plot results for the ALEs corresponding
to $D_{7}$ = 0, 0.1, and 0.2.
We see that the most severe experimental bounds on the abundance of
$^{7}$Li/H from Bonifacio {\it et al.}  agree with  the theoretical model for
$D_{7} \gtrsim 0.2$, even if we adopt the Low (D/H) constraint.

%%%%%%%%%%%%%%%%%%%%%%%%%%%%%%%%%%%%%%%%%%%%%%%%%%%%%%%%%%%%%%%%%%%%%%
%%%%%%%%%%%%%%%%%%%%%%%%%%%%%%%%%%%%%%%%%%%%%%%%%%%%%%%%%%%%%%%%%%%%%%

%%%%%%%%%%%%%%%%%%%%%%%%%%%%%%%%%%%%%%%%%%%%%%%%%%%%%%%%%%%%%%%%%%%%%%
\section{Discussion} \label{sec:discussion}
%%%%%%%%%%%%%%%%%%%%%%%%%%%%%%%%%%%%%%%%%%%%%%%%%%%%%%%%%%%%%%%%%%%%%%

We have investigated a scenario where late-decaying particles (LDPs)
of lifetime $\tau_X \gtrsim 1$\,sec, possessing a finite branching
ratio into hadrons, $B_{\rm h}$, and emitting visible particles of
energy $E_{\rm vis}$, are incorporated into the standard cosmological
model.  The scenario reaffirms the earlier prediction of DEHS that, as
now observed, the $^6$Li/$^7$Li abundance ratio is much larger than
the SBBN predicted value.  We find that this allows us to solve the
so-called ``Lithium problems'', while simultaneously remaining
consistent with all other observational constraints on the abundances
of light elements (ALEs) (within their 2 $\sigma$ error ranges).  In
other words, there exists a region of parameter space in which even
the most severe observational bounds can be satisfied.  It lies within
the range $1.5 \times 10^{-14} < B_{\rm h}(E_{\rm vis}/{\rm
1\,GeV})Y_{X} < 3.0 \times 10^{-13}$ (where again $Y_X$ is the ratio
of the primordial number density of LDPs divided by the entropy
density) and $3.0 < \log_{10}(\tau_X/{\rm 1\,sec})<3.8$ (with narrower
ranges corresponding to observational constraints that are {\it
stronger} than the most {\it mild} bounds discussed in this study).
We derived the allowed regions in terms of the generalised parameters
describing LDPs and converting them to some specific particle physics
model parameters should be straightforward. This also means that
the current study has independently confirmed the pioneering works
~\cite{Jedamzik:2004er, Jedamzik-Li} by Jedamzik and his collaborators, 
with an improved treatment of some of the key reactions involved, as 
discussed in \S,II.

The scenario has the possibility of the nonthermal-production of
cold/warm dark matter as one of the decay products of the $X$, which
must be important for formations of small scale
structures~\cite{WDM-SSS,Jedamzik-Li,Jedamzik:2007cp}.

Finally, since we have a lower limit on $B_{\rm h}E_{\rm vis}Y_X$ to
solve the Lithium problem, together with the constraint that the
additional component of the energy density of the parent massive
particle $X$ not to produce too much $^4$He, we can derive a lower
limit on the hadronic branching ratio as $B_{\rm h}>10^{-8}$.  The
precise constraint $B_h > 10^{-8}$ originates from the relatively
reasonable assumption that we allow one additional neutrino species
which contributes (approximately 15\% of the total) to the energy
density at the freezeout temperature of nucleons $T = 0.1$\,MeV .

%%%%%%%%%%%%%%%%%%%%%%%%%%%%%%%%%%%%%%%%%%%%%%%%%%%%%%%%%%%%%%%%%%%%%%
%\section{Acknowledgements}
%%%%%%%%%%%%%%%%%%%%%%%%%%%%%%%%%%%%%%%%%%%%%%%%%%%%%%%%%%%%%%%%%%%%%%

{\it Acknowledgements ---} We thank T. Moroi for valuable suggestions
and K. Jedamzik for useful comments. DTC is supported by a PPARC
scholarship. KK is supported in part by PPARC grant, PP/D000394/1, EU
grant MRTN-CT-2006-035863, the European Union through the Marie Curie
Research and Training Network "UniverseNet" and NASA grant
NNG04GL38G. GDS was supported by a grant from the US DOE, and by
fellowships from the John S. Guggenheim Memorial Foundation and the
Beecroft Institute for Particle Astrophysics and Cosmology at Oxford
which he thanks for hospitality.  This work was supported in part by
the Grant-in-Aid for Scientific Research from the Ministry of
Education, Science, Sports, and Culture of Japan, No.\,18540254 and
No.\,14102004 (M.K.) and No.\,18840010 (K.I.).  This work was also
supported in part by JSPS-AF Japan-Finland Bilateral Core Program
(M.K.).

\vspace{-0.5cm}

%%%%%%%%%%%%%%%%%%%%%%%%%%%%%%%%%%%%%%%%%%%%%%%%%%%%%%%%%%%%%%%%%%%%%%

\end{document}